# An Implementation of Partial Transmit Sequences to Design Energy Efficient Underwater Acoustic OFDM Communication System

**WALEED RAZA**[1,2,3], **XUEFEI MA**[1,2,3,4*], **AMIR ALI** [1,2,3], **ZUBAIR ALI SHAH**[5], **GHAZANFAR MEHDI**[5]

[1]College of Underwater Acoustic Engineering, Harbin Engineering University, Harbin 150001, China

[2]College of Engineering, Tibet University, Lhasa, 850000, China.

[3]Acoustic Science and Technology Laboratory, Harbin Engineering University, Harbin 150001, China;

[4]Key Laboratory of Marine Information Acquisition and Security (Harbin Engineering University), Ministry of Industry and Information Technology; Harbin 150001, China

[5]College of Power and Energy Engineering Harbin Engineering University, Harbin 150001, China

Corresponding author: Xuefei Ma (maxuefei@hrbeu.edu.cn).

This work was supported by the following projects: Equipment Prophetic Field Fund (No. 61404150301), Heilongjiang Natural Science Foundation Joint Guidance Project (No.LH2019A006), Equipment Pre-Study Ship Heavy Industry Joint Fund (No. 6141B042865), Underwater Information and Control Key Laboratory Open Fund (No. 6142218061812), Xiamen University Education Fund (No. UAC201804).

*waleed@hrbeu.edu.cn, amir@hrbeu.edu.cn, zubair_ali@hrbeu.edu.cn, ghazanfarmehdi22@gmail.com*

*Abstract*— **In this article we research about underwater acoustics transceivers. As Underwater acoustic transceivers consume more power than Radio frequency transceivers. The techniques which are being utilized in radio frequency cannot be implemented directly in underwater acoustic system it needs to be re investigated to design new methods. To achieve reliable acoustic data transmission new techniques should be achieved or the traditional Orthogonal frequency divisional multiplexing techniques should be revised. The power consumption also relies upon underwater acoustic signal propagation and transmission distances. Several underwater acoustic applications require long-term monitoring of the sea. For the battery powered modems, it becomes very serious problem. By designing an Energy efficient OFDM Communication system we can solve this problem. We study about peak to average power ratio in an Orthogonal frequency divisional multiplexing system by reducing the major draw-back of OFDM system. The PAPR reduction utilized in this paper is Partial Transmit Sequences for underwater acoustic OFDM communication system which has lesser complexity. The results have provided better performance in underwater acoustic OFDM communication system.**

## I. INTRODUCTION

Underwater acoustics sensing networks (UWASN) have the potential and ability to explore the ocean, Underwater acoustics networks can facilitate us in many applications. We can divide it into Civil and Military applications. Civil applications include such as oceanographic data collection, pollution monitoring, Offshore exploration, natural disaster prevention, assisted navigation and tactical surveillance[1]. Besides this there are many military applications such as for communication which meets the military needs. In the Battle field all the operations of the Navy rely on Underwater acoustics networks and communication. Underwater acoustic networking is promising technology and guaranties these applications. Furthermore, the commercial and civil applications are not limited here other application includes real-time pollution monitoring of streams, lakes, ocean bays, drinking water reservoirs, local ponds etc. [2]. Oil leakage detection equipment monitoring, and seismic imaging can be included. In addition, we can see biological behavior of different animals in the different in the different ocean. The Underwater acoustic communication (UAC) is key technique here to realize UWASN[2]. The underwater acoustic cable network has high cost and underwater cable deployment becomes very difficult, the need of sparse network topology and wireless communication is preferable for sensor communication [1]. The Radio frequency (RF) used in terrestrial networks can only propagate about several meters and the optical waves can only support underwater communication up to tens of meters. Magnetic coupled communication has been introduced in recent years for short range communication. By comparing all the wireless media, the underwater acoustic (UWA) signal has more significance, because it can propagate over several of kilometers (up to 40kilometers). This makes acoustic waves more suitable and attractive for communication. Therefore, Acoustic communication is key technology to realize the Underwater acoustic sensing networks UWASN[3, 4].

Acoustic signal propagation faces many challenges in underwater acoustic communication due to distinctive and nature of water environment at different ocean. For underwater acoustic signal propagation, the low frequency and distance dependent bandwidth is also a challenging task[3]. It limits the capacity and increases the transmit power in long distances communication. Due to multipath nature of underwater acoustic channel which results in large delay spreading causes inter symbol interference (ISI) and frequency selective fading[5], the propagation is much lower than Radio frequency (RF). The acoustic wave travels about at speed of 1500m/s. In addition, the environment of water for instance ocean waves, water temperature and salinity effects the propagation of underwater acoustic wave and induces high delay variance. The Large Doppler shifts is also a factor which is caused by ocean waves leads to vary the performance of underwater acoustic channel. Therefore, the network protocols designed for Radio frequency (RF) cannot be directly applied to Underwater acoustic systems[6]. New techniques should be achieved for reliable and efficient acoustic data transmissions. Each of above mentioned have their unique advantages by time. In FSK modulation scheme, the signal which has been transmitted it carries information bits to select the carrier frequencies. Afterwards measured power is compared by the receiver at several frequencies to conclude the transmission signal [10]. At the receiving side the energy detectors are used, this method bypasses the channel estimation. The interference caused by time spreading and frequency spreading is avoided with the help of guard intervals and guard band respectively. The waveform of narrow band W bandwidth is spread to a large bandwidth of B before transmission. Each symbol is multiplied with a spreading code of length N=[B/W][7]. De-spreading operation is used for multiple operation at receiving side, this process suppresses the induction caused by the time spreading. Before this spreading, to map information bits to symbols phase-coherent modulation is used. For single carrier modulation (PSK) phase-shift-key and quadrature-amplitude-modulation (QAM) constellations are used and it was first step towards high data rate communication[8, 9]. After sweep spread spectrum carrier modulation and we have multi-carrier modulation. The main theme of multi-carrier modulation dividing available bandwidth into many sub-bands, each of sub-bands have their own sub-carrier. In each of band symbol duration is increased for symbol rate, so that Inter-symbol interference can be less severe to reduce complexity of receiver and channel equalization[10-12].

Orthogonal frequency division multiplexing (OFDM) is an example of multicarrier modulation with overlapping subcarriers. To maintain orthogonality is key for the performance of the OFDM system. OFDM waveform is designed carefully over along multipath fading channel. It eliminates the need of an equalizer. In the recent decade OFDM technology has succeeded in broadband wireless radio applications, including broadcasting of audio/video, Wireless local area networks and all fourth-generation cellular networks. Along these application OFDM system have some severe drawbacks[4, 7, 11]. An OFDM waveform has defect of high PAPR. It restricts the linear dynamic range of transmitter's power amplifier and creates distortion which causes in the BER performance of the whole system. Due to Doppler spread it introduces significant interference among the OFDM subcarriers. Underwater acoustic channels have fast varying nature. It is difficult in this situation to implement same techniques of RF OFDM system in underwater acoustic system. The Signal processing specialized is needed here. Further identifying the PAPR problem in an OFDM system. The large superposition of data symbols on many subcarriers, an OFDM waveform has a large peak to average power ratio (PAPR). The Spectral efficiency of an OFDM system is also limited after the addition of Cyclic Prefix (CP) in every of OFDM symbol[13]. For long range underwater acoustic application, the PAPR is major concern because the power amplifier performs its full efficiency[14]. It needs very high-power scope to overcome the nonlinear distortion, it also requires a very high-power scope. The PAPR is different with the different occasions in the underwater acoustic sensing networks it relies on system operating conditions. For quantization of signal which has very high PAPR extra bits are needed here resulting it increases the complexity of transmitting transducer. Underwater acoustic (UWA) channels have large temporal variations, abundance of transmission paths, and wide property in nature. For underwater wireless communication system, acoustics waves area used as the primary carrier due to their relatively low absorption in underwater environment, the complexity of underwater acoustic transmission medium and low propagation speed of sound in water, regarded as the most challenging task for communication. Along with this challenge a wide range of underwater exploration and applications have emerged. The PAPR reduction techniques which are mainly considered in Underwater acoustic channels are coding technique, probability technique, and signal predistortion technique.

## II. RELATED WORK

Jinqiu wo and Gang qiao presented the work for reduction of PAPR in underwater acoustic OFDM system[15], which is based on amplitude limiting and improved companding transformation. However, companding is technique for reduction of peak to average power ratio. They did not provide the complexity of overall system. Xuefei ma researched about pulse shaping in GFDM system and its effects on PAPR performance in underwater acoustic 5G communication[16]. There focus was to implement the GFDM technique in underwater acoustic communication. In [17] Byung Moo Lee proposed energy efficient wireless communication system, he used selective mapping and clipping method to reduce the peak amplitude in an OFDM system, but it was employed in radio frequency communication. Menn S. Abd El-Galil researched the peak to average in underwater acoustic communication which relies on pre-equalization linear minimum mean square error (LMMSE) and zero forcing (ZF) pre-equalizers[18]. Jeong Wo han while designing a high speed underwater acoustic communication system used clipping reduction method to reduce the peak to average power ratio[19]. R.M Gomathi and J. Martin Leo presented DCT with LPDC codes to design the energy efficient underwater acoustic communication system which highly reduced the peak to average power ratio[20]. There are different approaches for reduction of PAPR proposed in the literature for terrestrial networks[10, 21-23].

In this research the energy efficient OFDM system is proposed in underwater acoustic sensing networks to handle this problem. The PAPR issue of OFDM modulation is briefly discussed and presented the effective PAPR reduction technique. We selected the parameters that are more suitable for underwater acoustic channel. Finally, the simulation results prove to be feasible for the underwater acoustic sensing networks. At the meantime we have used partial transmit sequence peak to average power ratio reduction technique to design low complexity system which is energy efficient and consumes less power for the power amplifier in Underwater acoustic sensor networks. we exhibit the PAPR issue of OFDM modulation, and then present the effective PAPR reduction techniques.

This Paper is designed as follows. In Section III the characteristics of OFDM signal is stated with the system model in underwater acoustic OFDM communication. The PAPR lowering strategies and PTS method is described briefly in sub section IV and the power efficiency thorough partial transmit sequence is derived. MATLAB simulation and numerical results are analyzed in section V Finally, conclusion and future work is shown in section VI.

## III. SYSTEM MODEL IN UNDERWATER ACOUSTIC OFDM COMMUNICATION

OFDM signal is formed of different subcarriers transmitting from transducer through modulation process in underwater acoustic communication. undergoing phase-shift keying (PSK) or quadrature amplitude modulation (QAM), constellation mapping can be obtained. In an OFDM system N represents the subcarriers, and input bit stream is shown as { $a_i$ }. The data passes serial to parallel conversion and constellation it form a new series of signal $\{d_0, d_1, \ldots d_i, \ldots d_{N-1}\}$ is obtained, $d_1$ can be regarded as the discrete complex values signal.[24]. If the value of, $d_i \in \{\pm 1\}$ to form a BPSK constellation. When QPSK mapping is adopted the value of $d_i \in \{\pm 1, \pm i\}$. The parallel signal sequence $e^{j2\pi f_0 t}, e^{j2\pi f_1 t}, \ldots e^{j2\pi f_{N-1} t}$ which is supplied to N orthogonal sub-carriers for the process of modulation [17]. In the last the OFDM symbols are formed by adding modulated signals together. Here we get advantage of discrete Fourier transform (DFT) which evaluates the performance of OFDM system design. After all these processes the OFDM signal is ready for transmission and complex envelope of the OFDM signal is as follows

$$x(t) = \frac{1}{\sqrt{N}} \sum_{k=0}^{N-1} X_k e^{j2\pi f_k t}, 0 \leq t \leq NT \quad (1)$$

The signals which have large number of subcarriers N can be regarded as Gaussian distribution with the function utilized probability density (PDF)

$$P_r\{x(t)\} = \frac{1}{\sqrt{2\pi\sigma}} e^{\frac{[x(t)]^2}{2\sigma^2}} \quad (2)$$

Where $\sigma$ is the variance of x(t).

The ratio between maximum instantaneous power and its power is called peak to average power ratio in an OFDM signal[25]. For underwater acoustic OFDM communication PAPR can be summarized in the following equation

$$X(n) = \frac{1}{\sqrt{N}} \sum_{K=0}^{N-1} d_k e^{j(2\pi nk/n)} \quad (3)$$

PAPR can be given by

$$PAPR[(xt)] = \frac{P_{PEAK}}{P_{AVERAGE}} = 10\log_{10} \frac{\max\left[|X(n)|^2\right]}{E\left[|x_n|^2\right]} \quad (4)$$

From the above equation the $P_{eak}$ shows the maximum output power and for average output power we represented with $A_{verage}$. The expected value of E[.] $x_n$ shows the real transmitted OFDM signal[10]. Figure 1 shows the framework of communication system in underwater acoustic OFDM. We apply IFFT (Inverse fast Fourier transform) operation here on modulated input symbols then $x_n$ can be written as under:

$$x_n = \frac{1}{\sqrt{N}} \sum_{K=0}^{N-1} X_k W_{N^{nk}} \quad (5)$$

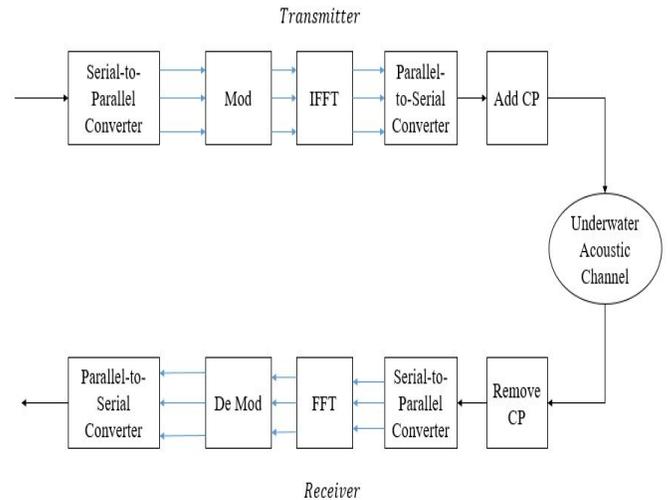

**Figure 01.** Block Diagram of underwater acoustic OFDM communication system

## IV. Partial Transmit Sequences and Power Efficiency

A lot of PAPR lowering techniques have been published in the Literature until today, but here we have used efficient use of PTS reduction technique which can be considered mainly in underwater acoustic channel.

Partial transmit sequence (PTS) method divides an input data block of N symbols into M disjoint sub blocks as defined below:

$$X = \left[X^0, X^1, X^2, \ldots, X^{M-1}\right]^T \quad (6)$$

here $X^i$ represents sub blocks which comes after one another and their size is same [21]. In every sub block scrambling is applied it means rotating the value of phase independently which is used in partial transmit sequences. After this process the separated sub blocks are multiplied with complex phase factor $b^m = e^{j\phi m}, m = 1, 2, \ldots, M,$ afterwards we apply the Inverse fast Fourier transform operation (IFFT) and will get:

$$x = IFFT\left\{\sum_{m=1}^{M} b^m X^m\right\} = \sum_{m=1}^{M} b^m .IFFT\{X^m\} = \sum_{m=1}^{M} b^m X^m \quad (7)$$

here $\{X^m\}$ is regarded as Partial Transmit Sequence (PTS)[21]. For minimum PAPR, phase factors are selected, Figure (2) shows the block diagram of partial transmit sequence.

$$[\tilde{b}^1,...\tilde{b}^M] = \arg\min_{[b^1,...,b^M]} \left(\max_{n=0,1,...,N-1} \left|\sum_{m=1}^{M} b^m x^m[n]\right|\right) \quad (8)$$

The minimum peak to average power ratio (PAPR) vector is defined mathematically in time domain as follows

$$\tilde{X} = \sum_{m=1}^{M} \tilde{b}^m X^m \quad (9)$$

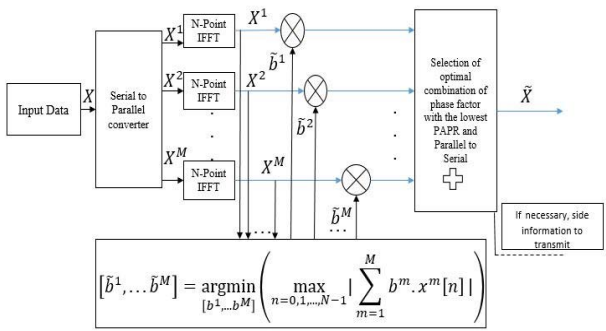

**Figure 02.** Layout of Partial Transmit Sequences in an OFDM system

Generally, for reduction of search complexity, the mathematical expression $\{b^m\}_{m=1}^{M}$ represents the phase factor and is restricted to a set of numbers. As allowed phase factor is $b = \{e^{j2\pi i/w} | i = 0,1,...,W-1\}, W^{M-1}$. To get the ideal set of phase vectors the set of all phase factor should be filtered. In this situation the search complexity increases greatly as the number of subblock increases. The $M$ Inverse fast Fourier transform (IFFT) operations is applied in each of data block as required by partial transmit sequences and $\lfloor \log_2 W^M \rfloor$ bits information which will be transmitted. The factors which affects the execution of partial transmit sequences are sub block partitioning, number of subblocks $M$ and phase factors which are allowed $W$. the sub block partitioning furthermore is divided into three types the first one is Adjacent, and the second category is Interleaved, and third type is pseudo-random. All three have better performance but among all pseudo-random has provided the better performance as we discussed earlier. However, the search complexity is increased in this PAPR reduction method when the number of subblock increases the complexity also becomes too high. In this paper we used Adjacent sub block partitioning in partial transmit sequence to reduce the PAPR, by using binary phase factor [26] which is related to suboptimal algorithm.

Further evaluating the performance of PTS reduction technique is sub-block partitioning. In this method the subcarriers are divided into many disjoint sub-blocks. Adjacent, Interleaved, and Pseudo-random partitioning are three types of this partitioning. The PTS technique is based on an arbitrary number of subcarriers and all types of modulation. In a distribution adjacent each sub-block to a regular structure containing approximately D / M the active carriers and the rest of the non-active positions filled with zeros. The sub-blocks represent the carriers so that a carrier can only be represented once in a set of sub-blocks. In occurrence D represents the sub-carrier number and M shows the number of sub-blocks. A pseudo-random distribution has no regular structure for the allocation of carriers to the sub- blocks unlike an Interleaved distribution which means that each sub-block follows a regular structure with an active carrier followed by M-1 zeros. The other sub-blocks represent an offset version of this structure[23].

The example of PTS technique is shown here for OFDM system with number of subcarriers is eight then the subcarriers are further divided into 4 sub-blocks. We select the phase factor at P = {± 1}.

The distribution of adjacent sub-block can be seen in figure (3) when a data block of X length = 8. There are eight 8 (= $2^{4-1}$) ways to merge the sub-blocks by setting $b_1$= 1. Among them [b1, b2, b3, b4]$^T$= [1, -1, -1, -1]$^T$ makes it possible to obtain the minimum PAPR.

After all this the transformed data block will be:

$$\sum_{m=1}^{M} b_m X_m = [1,-1,-1,1,-1,1,1,1]^T.$$

The number of IFFT operations are 4 which are required for this case and the amount of lateral information is 4. To retrieve the original data block, this side information must be transmitted to receiver. There are different ways to send this information one of them is to send this data with a separate channel which is different from the data channel. The Lateral information can be included in the data block; however, this gives output in loss of data throughput.

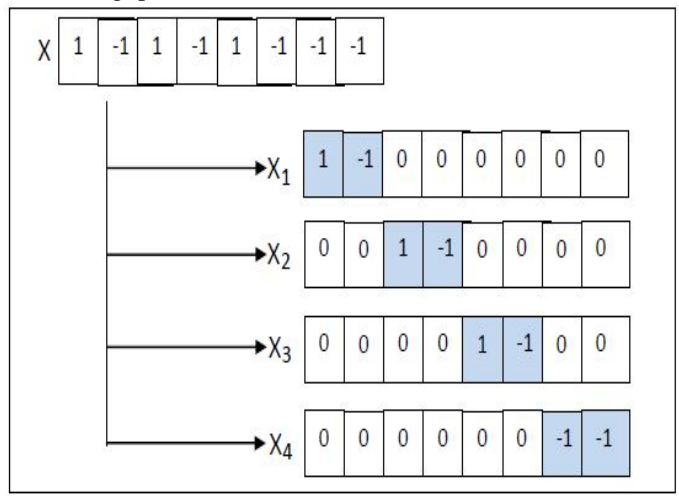

**Figure 03.** An example of an adjacent PTS distribution with 8 subcarriers in 4 sub-blocks

### A. Power Efficiency

The average input should be adjusted in an OFDM system to decrease the distortion effects in peak signal, here input back off (IBO) is required for implementation. IBO is used as performance metric for reduction of the input power and required input power can be calculated. So, we can get estimated output power. This is applied because an IBO is directly connected with PAPR as well as the efficiency $\eta$, the IBO and efficiency has indirect relation when the value of IBO higher it means efficiency $\eta$ is lesser, sometimes an IBO is equivalent to PAPR. Power amplifier which are employed in an OFDM system their efficiency is given as under

$$\eta = \frac{P_{out,avg}}{P_{DC}} \quad (10)$$

As we know there are different kinds of amplifiers. Class A amplifiers have less efficiency as compared to linear amplifier. Class A amplifiers efficiency lies in between 10-25% but not more than 50%. Hence, the saturation point is maintained in an ideal amplifier. The efficiency can be shown mathematically as under

$$\eta = \frac{0.5}{PAR} \quad (11)$$

Power consumption $P_{DC}$ can be regarded as the power saving for the system

$$P_{DC} = \frac{P_{out,avg}}{\eta} \quad (12)$$

Now, by putting the equation (11) into (12), the equation we will get is

$$P_{DC} = \frac{P_{out,avg}}{\frac{1}{2PAR}} \quad (13)$$

$$P_{DC} = 2P_{out,avg} PAR \quad (14)$$

Hence another form of power efficiency can be written as follows in terms of power saving

$$P_{savings} = 2P_{out,ave}(PAR_{initial} - PAR_{final}) \quad (15)$$

For calculating power savings gain $G_s$, we will show the saving gain $G_s$ as

$$G_s = \frac{P_{savings}}{P_{out,avg}} \quad (16)$$

By substituting the equation (15) into (16) we will get

$$G_s = \frac{2P_{out,ave}(PAR_{initial} - PAR_{final})}{P_{out,ave}} \quad (17)$$

Therefore, for the PAPR saving gain $G_s$ is defined as

$$G_s = 2(PAR_{initial} - PAR_{final}) \quad (18)$$

### V. Simulation & Numerical Analysis

To authenticate the feasibility of our novel technique, simulutions were conducted in MATLAB. In this simulation the reduction performance of PAPR is simulated based on signal scrambling in an OFDM symbol with the help of complementary cumulative distribution function (CCDF). For the proposed novel scheme we put the parameters that are more suitable for underwater acoustic channel.

| NO. | Parameter | Data |
|---|---|---|
| 1 | Sampling frequency | 100k |
| 2 | Bandwidth | 6.25k |
| 3 | FFT points | 8192 |
| 4 | OFDM symbol number | 23 |
| 5 | Number of subcarriers | 1024 |
| 6 | Cyclic prefix time | 25ms |
| 7 | Modulation | QPSK |
| 8 | Sound speed | 1500m/s |

The number of subcarriers were kept 1024 in an OFDM signal with QAM constellation the signal was oversampled with L=4.The number of sub blocks were kept M=16 employing random phase factors used in PTS technique. The results obtained after simulation are shown in the Figure 4, 5, 6 and 7 respectively

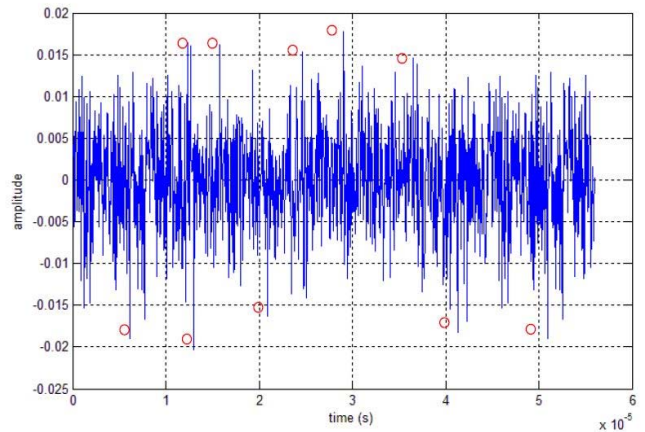

**Figure 04.** The Simulated Time Domain OFDM Signal

Figure 4 illustrates OFDM signal in the time domain after simulation. We can observe very clearly that among several drawbacks of an OFDM signal it has high peak to average power ratio the existence of extremely high peaks, it is the major disadvantage of the OFDM signals.

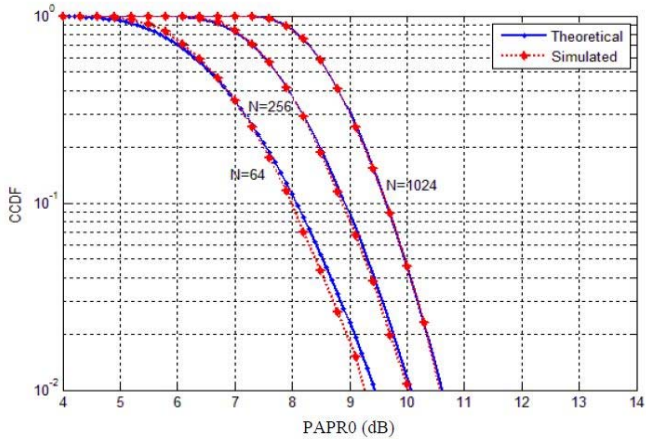

**Figure 05.** CCDF of OFDM Signal with different numbers of subcarriers (N)

Figure 5 shows the probability of occurrence of high levels of PAPR of a discreet time and baseband OFDM signal knowing the number of subcarriers constituent. The theoretical equation is confronted with the results obtained by simulation to confirm its correct interpretation. As the result suggests theoretical, the occurrence of high levels of PAPR increases rapidly with the increment of subcarriers used. When the subcarriers are high the value of PAPR also increases. In the simulation note the number of subcarriers varies between 64, 256 and 1024. It can be clearly seen that the value of PAPR becomes enlarged with the increment in the sub-carriers. As it is shown in the figure, the PAPR is 9.3 dB for 64 subcarriers, for the 256 it is 10 dB and finally for 1024 subcarriers we will have a PAPR of 10.6 dB.

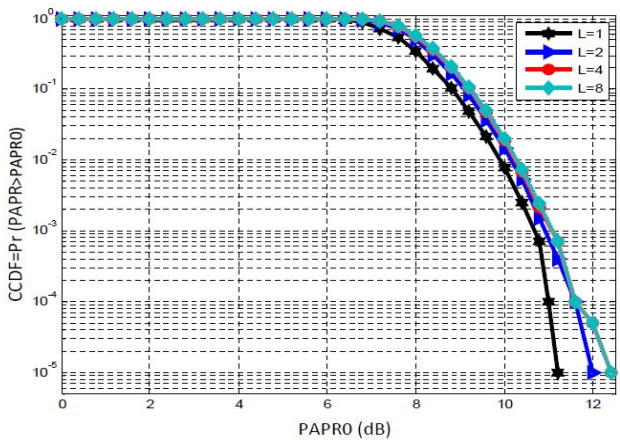

**Figure 06.** Complementary Distribution Function of the OFDM signal when N = 256 carriers varying the value of Oversampling.

Figure 6 shows the probability of occurrence high levels of PAPR for oversampled OFDM signals. We note that for values of oversampling factors L > 4, the probability of exceeding PAPR does not present any significant evolution. The curves do not change further. We will therefore use an oversampling of L = 4 to approximate signals OFDM.

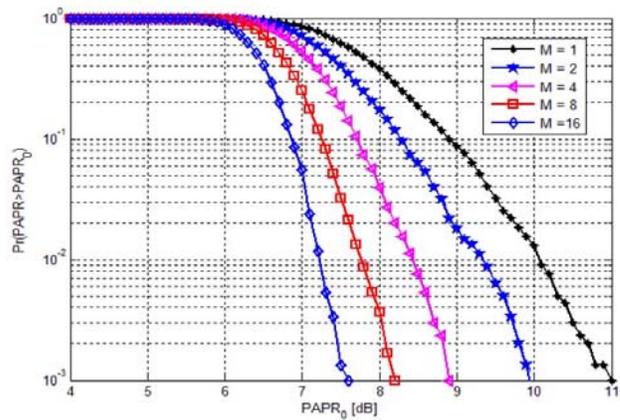

**Figure 07.** Different number of subblocks in Partial Transmit Sequence

Figure 7 illustrates that the performance of PTS is dependent of number of subblocks M, by increasing the value of sub blocks the method becomes more efficient. On the other hand, when the value of subblock is higher, the higher the number of combinations. It means more increase and comparisons, resulting in more time for processing and more memory.

The following table (2) shows the complexity of the system with the increase in the number of under block M by calculating the possibilities obtained from the candidate sequences to optimize at the exit PTS block. The calculation is done by the Formula $W^{(M-1)}$ if one takes W = 4 {number of factor phase $1,-1,j,-j$} we find:

**Table 2.** Complexity of OFDM System with Partial Transmit Sequence.

| M | 1 | 2 | 4 | 8 | 16 |
|---|---|---|---|---|---|
| Number of Opportunities Optimized | 1 | 4 | 64 | 16484 | 1073741824 |

To verify the performance of partial transmit sequences and power saving through above equations we conducted simulations in MATLAB to verify the feasibility of mentioned equation. Figure (7) illustrates the performance of PAPR of underwater acoustic OFDM system using bell hop multipath channel. The higher PAPR reduction is pursued by enlarging sub-blocks M in Partial Transmit Sequences as in figure. Our research aims on power saving of underwater acoustic OFDM modem through Partial Transmit Sequences. We discussed previously that the saving gain is difference in PAPR $G_s = 2(PAR_{initial} - PAR_{final})$. The overview of several PAPR reduction performances and measurement can be seen in Table 3, these values correspondence to the curves. By enlarging partial transmit sequence sub-blocks saving gain can be achieved. From equation (18) shows increasing in saving gain, the power saving will increase. Hence, we can achieve power saving by PTS.

**Table 3.** PAPR of OFDM signal & Saving Gain $G_s$ at different number of Subblocks.

| Number of Sub-blocks(M) No. of PTS | PAPR Peak to Average Power Ratio(dB) | $G_s$ (Saving Gain) |
|---|---|---|
| 1 | 11 | 0 |
| 2 | 9.9 | 4.1 |
| 4 | 8.88 | 5.8 |
| 8 | 8.25 | 6.3 |
| 16 | 7.55 | 7.8 |

The reduction performance of the PAPR is dependent of the number of sub-blocks M and the value of allowed phase factors W. the greater the number of subblocks increases the complexity but improves the performance. As well as, the higher the number as the number of phase factors increases, the complexity can become too high. Simulation results and numerical analysis proves that this novel design of PTS Technique can be used over underwater acoustic channel with efficient battery power usage in underwater acoustic networks. As the selection of minimum operation occurs in this technique the high peaks can be removed in an OFDM system. In this way, the overall system will work efficiently, and it will consume less power for the power amplifier in battery powered transceivers.

## VI. Conclusion and Future Work

In this paper we have measured the peak to average power ratio of underwater acoustic OFDM communication system to design a novel an energy efficient OFDM system for underwater acoustic network. This system is very useful in future for long term battery deployed application. According to environment of underwater acoustic channel conditions, we need to choose a suitable method due to the severe surroundings and limited bandwidth. The proposed technique is highly energy efficient and have less computational complexity. From the results it can be concluded that the proposed Partial Transmit sequence method gives the better outcome in terms of PAPR reduction performance. We plan to continue this work in the future by introducing some innovative less complexity methods for reduction of the PAPR in underwater acoustic OFDM communication.

**RAZA WALEED** received the B.E degree in Electronic Engineering from the Department of Electronic Engineering, Dawood University of Engineering and Technology Karachi, Pakistan in 2017.He is currently pursuing the M.S. degree in underwater acoustic communication engineering in the College of Underwater Acoustic Engineering, Harbin Engineering University, Harbin China. His research area of interests includes underwater acoustic communication.

**XUEFEI MA** received the B.S. degree in electronic information engineering from the Harbin Engineering University, China, in 2003, and the M.S. degree in information and communication engineering from the Harbin Engineering University, China, in 2006,and the Ph.D. degree in Signal and information processing engineering from the School of College of Underwater Acoustic Engineering, Harbin Engineering University, China, in 2011, where he is currently an Associate Professor. His research interests include the underwater acoustic detection, underwater acoustic communication, and underwater acoustic confrontation.

**ALI AMIR** received the B.E degree in Electronic Engineering from Mehran University of Engineering Technology Jamshoro, Pakistan in 2016. Currently he is getting M.S. degree from the Harbin Engineering University, China. His current research interests include system engineering, underwater signal processing and underwater acoustic communication.

**ZUBAIR AI SHAH** received the B.E. degree in Mechanical Engineering, from Mehran University of Engineering Technology Jamshoro, Pakistan in 2016. He is currently pursuing MS degree in Power and Energy Engineering from the Harbin Engineering University, China. His current research interests include organic Rankine recycle in thermo-physics.

**GHAZANFAR MEHDI** received the B.E. degree in Mechanical Engineering from Quaid-Awam University Nawab-Shah Pakistan, in 2016 and MS degree in power and energy engineering from Harbin Engineering University China 2019. He is currently pursuing the PHD degree in Mechanical Engineering from University of Salento, Italy. His research interests include the renewable energy and thermo-physics.

References:
[1] Z. W. Shengli Zhou, *OFDM for Underwater Acoustic Communications* (Communication Technology). United Kingdom: John Wiley & Sons, Ltd., June 2014, p. 410.
[2] T. B. Santoso, Wirawan, and G. Hendrantoro, "Development of underwater acoustic communication model: Opportunities and challenges," presented at the 2013 International Conference of Information and Communication Technology (ICoICT), 2013.
[3] D. Lucani, M. Medard, and M. Stojanovic, "Underwater Acoustic Networks: Channel Models and Network Coding Based Lower Bound to Transmission Power for Multicast," *IEEE Journal on Selected Areas in Communications,* vol. 26, no. 9, pp. 1708-1719, 2008, doi: 10.1109/jsac.2008.081210.
[4] R. P. I. MEMBER. High Rate OFDM Acoustic Link for Underwater Communication
[5] M. S. a. a. J. G. P. Ethem M. Sozer, "<Underwater acoustic Network.pdf>," *IEEE JOURNAL OF OCEANIC ENGINEERING,* vol. VOL. 25, 2000.

[6]     R. Ashri, H. Shaban, and M. El-Nasr, "A Novel Fractional Fourier Transform-Based ASK-OFDM System for Underwater Acoustic Communications," *Applied Sciences,* vol. 7, no. 12, 2017, doi: 10.3390/app7121286.

[7]     X. Wang, X. Wang, R. Jiang, W. Wang, Q. Chen, and X. Wang, "Channel Modelling and Estimation for Shallow Underwater Acoustic OFDM Communication via Simulation Platform," *Applied Sciences,* vol. 9, no. 3, 2019, doi: 10.3390/app9030447.

[8]     C. C. Jurong Bai, Yi Yang Feng, Zhao Xiangjun, Xin Abdel-Hamid, Soliman Jiamin Gong, "Peak-to-average power ratio reduction for DCO-OFDM underwater optical wireless communication system based on an interleaving technique," *Optical Engineering,* vol. Paper 180820, 2018, doi: 10.1117/1.OE.57.8.086110.

[9]     D. R. K M, S.-H. Yum, E. Ko, S.-Y. Shin, J.-I. Namgung, and S.-H. Park, "Multi-Media and Multi-Band Based Adaptation Layer Techniques for Underwater Sensor Networks," *Applied Sciences,* vol. 9, no. 15, 2019, doi: 10.3390/app9153187.

[10]    L. Hao, D. Wang, Y. Tao, W. Cheng, J. Li, and Z. Liu, "The Extended SLM Combined Autoencoder of the PAPR Reduction Scheme in DCO-OFDM Systems," *Applied Sciences,* vol. 9, no. 5, 2019, doi: 10.3390/app9050852.

[11]    J. Wu, "Iterative Compressive Sensing for the Cancellation of Clipping Noise in Underwater Acoustic OFDM System," *Wireless Personal Communications,* vol. 103, no. 3, pp. 2093-2107, 2018, doi: 10.1007/s11277-018-5897-9.

[12]    G. Wunder, R. F. H. Fischer, H. Boche, S. Litsyn, and J.-S. No, "The PAPR Problem in OFDM Transmission: New Directions for a Long-Lasting Problem," *IEEE Signal Processing Magazine,* vol. 30, no. 6, pp. 130-144, 2013, doi: 10.1109/msp.2012.2218138.

[13]    S. Xing, G. Qiao, and L. Ma, "A Blind Side Information Detection Method for Partial Transmitted Sequence Peak-to-Average Power Reduction Scheme in OFDM Underwater Acoustic Communication System," *IEEE Access,* vol. 6, pp. 24128-24136, 2018, doi: 10.1109/access.2018.2829620.

[14]    B. Lee and Y. Kim, "Transmission Power Determination Based on Power Amplifier Operations in Large-Scale MIMO-OFDM Systems," *Applied Sciences,* vol. 7, no. 7, 2017, doi: 10.3390/app7070709.

[15]    G. Q. Jinqiu Wu, and Xiaofei Qi, "The Research on Improved Companding Transformation for Reducing PAPR in Underwater Acoustic OFDM Communication System," *Discrete Dynamics in Nature and Society,* vol. Article ID 3167483,, 2016, doi: 10.1155/2016/3167483.

[16]    X. M. Jinqiu Wu, Xiaofei Qi, Zeeshan Babar, and Wenting Zheng, "Influence of Pulse Shaping Filters on PAPR Performance of Underwater 5G Communication System Technique: GFDM," *Wireless Communications and Mobile Computing,* vol. Volume 2017, Article ID 4361589, 2017, doi: 10.1155/2017/4361589.

[17]    B. M. Lee, Y. S. Rim, and W. Noh, "A combination of selected mapping and clipping to increase energy efficiency of OFDM systems," *PLoS One,* vol. 12, no. 10, p. e0185965, 2017, doi: 10.1371/journal.pone.0185965 A combination of selected mapping and clipping to increase energy efficiency of OFDM systems

[18]    M. S. Abd El-Galil, N. F. Soliman, M. I. Abdalla, and F. E. Abd El-Samie, "Efficient underwater acoustic communication with peak-to-average power ratio reduction and channel equalization," *International Journal of Speech Technology,* vol. 22, no. 3, pp. 649-696, 2019, doi: 10.1007/s10772-019-09600-1.

[19]    J.-w. Han, S.-y. Kim, K.-m. Kim, S.-y. Chun, and K. Son, "Design of OFDM System for High Speed Underwater Communication," presented at the 2009 International Conference on Computational Science and Engineering, 2009.

[20]    R. M. G. a. J. M. L. Manickam, "PAPR REDUCTION TECHNIQUE USING COMBINED DCT AND LDPC BASED OFDM SYSTEM FOR UNDERWATER ACOUSTIC COMMUNICATION," *ARPN Journal of Engineering and Applied Sciences,* vol. VOL. 11, 2016.

[21]    A. S. Amhaimar Lachen, Asselman Adel, "<Low-Computational-Complexity-PTS-Scheme-for-PAPR-Reduction," *Procedia Engineering 181 ( 2017 ) 876 – 883,* 2017, doi: 10.1016/j.proeng.2017.02.480.

[22]    L. R. Luo Renze, Dang Yupu,Yang Jiao & Liu Weihong, "Two improved SLM methods for PAPR and BER reduction in OFDM–ROF systems," *Optical Fiber Technology,* vol. R. Luo et al. / Optical Fiber Technology 21 (2015) 26–33, 2014, doi: 10.1016/j.yofte.2014.07.007.

[23]    M. K. Manju Bala, Kirti Rohilla, "Low-Complexity PAPR Reduction of OFDM Signals Using Partial Transmit Sequence (PTS)," *International Journal of Engineering and Innovative Technology (IJEIT),* vol. Volume 3,, no. Issue 11, 2014.

[24]    P. P. B. Suverna Sengar, "PERFORMANCE IMPROVEMENT IN OFDM SYSTEM BY PAPR REDUCTION," *Signal & Image Processing : An International Journal (SIPIJ,* vol. Vol.3, 2012, doi: 10.5121/sipij.2012.3211.

[25]    !!! INVALID CITATION !!! [17, 18].

[26]    K. Bakht *et al.*, "Power Allocation and User Assignment Scheme for beyond 5G Heterogeneous Networks," *Wireless Communications and Mobile Computing,* vol. 2019, pp. 1-11, 2019, doi: 10.1155/2019/2472783.